\documentclass{pasj01}

\begin{document}

\author{Mariko~\textsc{Kimura},\altaffilmark{1,2,*}
        Keisuke~\textsc{Isogai},\altaffilmark{1,3}
        Taichi~\textsc{Kato},\altaffilmark{1}
        Naoto~\textsc{Kojiguchi},\altaffilmark{1}
        Yasuyuki~\textsc{Wakamatsu}, \altaffilmark{1}
        Ryuhei~\textsc{Ohnishi},\altaffilmark{1}
        Yuki~\textsc{Sugiura},\altaffilmark{4}
        Hanami~\textsc{Matsumoto},\altaffilmark{4}
        Sho~\textsc{Sumiya},\altaffilmark{4}
        Daiki~\textsc{Ito},\altaffilmark{4}
        Kengo~\textsc{Nikai},\altaffilmark{4}
        Katsura~\textsc{Matsumoto},\altaffilmark{4}
        Sergey Yu.~\textsc{Shugarov},\altaffilmark{5,6}
        Natalia~\textsc{Katysheva},\altaffilmark{5}
        Hiroshi~\textsc{Itoh},\altaffilmark{7}
        Pavol A.~\textsc{Dubovsky},\altaffilmark{8}
        Igor~\textsc{Kudzej},\altaffilmark{8}
        Hiroshi~\textsc{Akitaya},\altaffilmark{9,12}
        Kohei~\textsc{Oide},\altaffilmark{10}
        Takahiro~\textsc{Kanai},\altaffilmark{10}
        Chihiro~\textsc{Ishioka},\altaffilmark{10}
        Yumiko~\textsc{Oasa},\altaffilmark{9,10,11}
        Tonny~\textsc{Vanmunster},\altaffilmark{13}
        Arto~\textsc{Oksanen},\altaffilmark{14}
        Tam\'{a}s~\textsc{Tordai},\altaffilmark{15}
        Katsuhiro L.~\textsc{Murata},\altaffilmark{16}
        Kazuki~\textsc{Shiraishi},\altaffilmark{16}
        Ryo~\textsc{Adachi},\altaffilmark{16}
        Motoki~\textsc{Oeda},\altaffilmark{16}
        Yutaro~\textsc{Tachibana},\altaffilmark{16}
        Seiichiro~\textsc{Kiyota},\altaffilmark{17}
        Elena P.~\textsc{Pavlenko},\altaffilmark{18}
        Kirill~\textsc{Antonyuk},\altaffilmark{18}
        Oksana~\textsc{Antonyuk},\altaffilmark{18} 
        Nikolai~\textsc{Pit},\altaffilmark{18}
        Aleksei~\textsc{Sosnovskij},\altaffilmark{18}
        Julia~\textsc{Babina},\altaffilmark{18}
        Alex~\textsc{Baklanov},\altaffilmark{18}
        Koji S.~\textsc{Kawabata},\altaffilmark{12}
        Miho~\textsc{Kawabata},\altaffilmark{12}
        Tatsuya~\textsc{Nakaoka},\altaffilmark{12}
        Masayuki~\textsc{Yamanaka},\altaffilmark{12}
        Kiyoshi~\textsc{Kasai},\altaffilmark{19}
        Ian~\textsc{Miller},\altaffilmark{20}
        Stephen M.~\textsc{Brincat},\altaffilmark{21}
        Wei~\textsc{Liu},\altaffilmark{22}
        Mahito~\textsc{Sasada},\altaffilmark{12}
        Daisaku~\textsc{Nogami}\altaffilmark{1}
        }
\email{mariko.kimura@riken.jp}

\altaffiltext{1}{Department of Astronomy, Graduate School of Science, Kyoto University, Oiwakecho, Kitashirakawa, Sakyo-ku, Kyoto 606-8502}

\altaffiltext{2}{Extreme Natural Phenomena RIKEN Hakubi Research Team, Cluster for Pioneering Research, RIKEN, 2-1 Hirosawa, Wako, Saitama 351-0198}

\altaffiltext{3}{Okayama Observatory, Kyoto University, 3037-5 Honjo, Kamogatacho, Asakuchi, Okayama 719-0232}

\altaffiltext{4}{Osaka Kyoiku University, 4-698-1 Asahigaoka, Kashiwara, Osaka 582-8582}

\altaffiltext{5}{Sternberg Astronomical Institute, Lomonosov Moscow State University, Universitetsky Ave., 13, Moscow 119992, Russia}

\altaffiltext{6}{Astronomical Institute of the Slovak Academy of Sciences, 05960 Tatranska Lomnica, Slovakia}

\altaffiltext{7}{Variable Star Observers League in Japan (VSOLJ), 1001-105 Nishiterakata, Hachioji, Tokyo 192-0153}

\altaffiltext{8}{Vihorlat Observatory, Mierova 4, Humenne, Slovakia}

\altaffiltext{9}{Graduate School of Science and Engineering, Saitama University, 255 Shimo-Okubo, Sakura-ku, Saitama, 338-8570}

\altaffiltext{10}{Graduate school of Education, Saitama University, 255 Shimo-Okubo, Sakura-ku, Saitama, 338-8570}

\altaffiltext{11}{Faculty of Education, Saitama University, 255 Shimo-Okubo, Sakura-ku, Saitama, 338-8570}

\altaffiltext{12}{Hiroshima Astrophysical Science Center, Hiroshima University, Higashi-
Hiroshima, Hiroshima 739-8526}

\altaffiltext{13}{Center for Backyard Astrophysics (Belgium), Walhostraat 1A, B-3401, Landen, Belgium}

\altaffiltext{14}{Hankasalmi observatory, Jyvaskylan Sirius ry, Verkkoniementie 30, FI-40950 Muurame, Finland}

\altaffiltext{15}{Polaris Observatory, Hungarian Astronomical Association, Laborc utca 2/c, 1037 Budapest, Hungary}

\altaffiltext{16}{Department of Physics, Tokyo Institute of Technology, 2-12-1
Ookayama, Meguro-ku, Tokyo 152-8551}

\altaffiltext{17}{VSOLJ, 7-1 Kitahatsutomi, Kamagaya, Chiba 273-0126}

\altaffiltext{18}{Federal State Budget Scientific Institution ``Crimean Astrophysical Observatory of RAS'', Nauchny, 298409, Republic of Crimea}

\altaffiltext{19}{Baselstrasse 133D, CH-4132 Muttenz, Switzerland}

\altaffiltext{20}{Furzehill House, Ilston, Swansea, SA2 7LE, UK}

\altaffiltext{21}{Flarestar Observatory, San Gwann SGN 3160, Malta}

\altaffiltext{22}{Purple Mountain Observatory, Chinese Academy of Sciences, No.10 Yuanhua Road, Qixia District, Nanjing, 210023, China}

\title{Multi-wavelength photometry during the 2018 superoutburst of the WZ Sge-type dwarf nova EG Cancri}

\Received{} \Accepted{}


\SetRunningHead{Kimura et al.}{The 2018 superoutburst of EG Cnc}

\maketitle

\begin{abstract}

We report on the multi-wavelength photometry of the 2018 
superoutburst in EG Cnc.  
We have detected stage A superhumps and long-lasting late-stage 
superhumps via the optical photometry and have constrained 
the binary mass ratio and its possible range.  
The median value of the mass ratio is 0.048 and the upper limit 
is 0.057, which still implies that EG Cnc is one of the possible 
candidates for the period bouncer.  
This object also showed multiple rebrightenings in this superoutburst, 
which are the same as those in its previous superoutburst 
in 1996--1997 despite the difference in the main superoutburst.  
This would represent that the rebrightening type is inherent to 
each object and is independent of the initial disk mass 
at the beginning of superoutbursts.  
We also found that $B-I$ and $J-K_{\rm S}$ colors were 
unusually red just before the rebrightening phase and became bluer 
during the quiescence between rebrightenings, which would mean 
that the low-temperature mass reservoir at the outermost disk 
accreted with time after the main superoutburst.  
Also, the ultraviolet flux was sensitive to rebrightenings 
as well as the optical flux, and the $U-B$ color became redder 
during the rebrightening phase, which would indicate that 
the inner disk became cooler when this object repeated 
rebrightenings.  
Our results thus basically support the idea that the cool 
mass reservoir in the outermost disk is responsible for 
rebrightenings.  

\end{abstract}

\section{Introduction}

Cataclysmic variables (CVs) are close binary systems consisting 
of a white dwarf (the primary star) and a low-mass star 
(the secondary star).  An accretion disk is formed around 
the primary via Roche-lobe overflow from the secondary.  
Dwarf novae (DNe) are a subclass of CVs and show outbursts 
representing a sudden release of gravitational energy.  
The mass accumulated at the disk in the quiescent state 
accretes onto the primary in the outburst state and 
a huge amount of gravitational energy is released during 
a short duration.  
This sudden accretion is triggered by the thermal-viscous instability 
in the disk (see \cite{war95book} and \cite{osa96review} for 
comprehensive reviews).   

SU UMa-type stars, one of the subclasses of DNe, have short orbital 
periods ($\sim$1 hr $<$ $P_{\rm orb}$ $<$ $\sim$ 2 hr) and show 
occasional superoutbursts which are defined as large-amplitude 
($\sim$6--8 mag) outbursts with superhumps.  
The superhumps are small-amplitude periodic light variations 
with periods slightly longer than the orbital period and 
are believed to be induced by the 3:1 resonance tidal instability 
\citep{osa89suuma,whi88tidal,hir90SHexcess,lub91SHa,lub91SHb}.
\citet{Pdot} investigated the time evolution of superhumps and 
proved that the superhumps have three stages: 
stage A superhumps with a long and constant period and increasing 
amplitudes, stage B superhumps with a systematically varying period 
and decreasing amplitudes, and stage C superhumps with a short and 
constant period.  

WZ Sge-type stars, a subclass of SU UMa-type stars, have 
extremely small binary mass ratios and predominantly show 
superoutbursts.  
Here the mass ratio is defined as the ratio of the secondary mass 
with respect to the primary mass ($q \equiv M_2/M_1$).  
Double-peaked modulations called early superhumps and rebrightening 
are their two representative observational properties 
(see \cite{kat15wzsge} and references therein).  
Early superhumps have a period almost equal to the orbital one 
and are observed at the early stage of superoutbursts 
\citep{kat02wzsgeESH,ish02wzsgeletter}.
They are considered to be triggered by the 2:1 resonance tidal 
instability \citep{osa02wzsgehump,osa03DNoutburst}.  
Rebrightening is usually observed soon after the main superoutburst 
with a long plateau stage and is classified into five types 
according to the morphology of light curves: 
type-A (long duration rebrightening), type-B (multiple rebrightenings), 
type-C (single rebrightening), type-D (no rebrightening), and 
type-E (double plateaus) \citep{ima06tss0222,Pdot,Pdot5}.  

Rebrightenings in WZ Sge stars are not yet explained well 
by the current model of the disk instability.  
Some people argued that the mass transfer from 
the secondary star is temporally enhanced via 
the irradiation by the hot accretion disk 
\citep{ham00DNirradiationreview,pat02wzsge}.  
However, it would be difficult for the irradiation to raise 
the mass transfer rate \citep{osa04EMT}, and no observational 
evidence of the increase of the mass-transfer rate has been 
detected.  
On the other hand, there is a possibility that 
substantial mass remains outside the 3:1 resonance radius, 
which accretes on the white dwarf delayed against 
the main superoutburst \citep{kat98super,hel01eruma}.  
A large amount of mass seems to be accumulated in the disk 
before superoutbursts because of long quiescence in WZ Sge stars, 
and once a superoutburst is triggered, the disk rapidly expands 
due to the angular-momentum transfer in the disk.  
Then the substantial mass is possibly conveyed beyond the 3:1 
resonance radius \citep{osa01egcnc}.  
The observational features like the presence of superhumps, 
the strong NaD absorption, the unusually red color, and 
the infrared activities, which were confirmed after 
the main superoutburst in some WZ Sge stars may support 
this idea 
\citep{kat97egcnc,pat98egcnc,uem08j1021,iso15ezlyn,nak14j0754j2304,neu17j1222}.  
Moreover, TCP J21040470$+$4631129, a recently observed 
WZ Sge-type star, triggered superoutbursts twice 
as a part of multiple rebrightenings after the main 
superoutburst \citep{tam20j2104}.  
This implies that there was a large amount of mass reservoir 
in the outermost part of the disk and that the enhanced 
mass transfer making the disk shrink unlikely occurred.  

WZ Sge stars are interesting targets also in the study of 
the CV evolution because some of them are believed to be 
the best candidates for period bouncers.  
Period bouncers are defined as the post-period-minimum CVs 
containing degenerate secondaries.  
CVs evolve as the orbital period becomes shorter for a long time 
due to angular momentum losses by magnetic breaking and 
gravitational-wave radiation.  
Once the secondary star becomes degenerate, the system evolves as 
the orbital period becomes longer, since the thermal timescale of 
the secondary becomes longer than the mass-transfer timescale 
(see \cite{kni11CVdonor} and references therein).  
Although the population of period bouncers is predicted to be 
$\sim$70\% of CVs by theoretical works, the number of 
the good candidates identified by observations is only around 
a few dozen (e.g., \cite{kol93CVpopulation,lit08eclCV,nak14j0754j2304}).  
The gap between observational and theoretical populations of 
period bouncers is one of the big problems in the CV evolution.  

\citet{kat13qfromstageA} proposed a new dynamical method for 
estimating the binary mass ratio by using the period of 
stage A superhumps.  The new method was applied to many 
WZ Sge-type stars, resulting in the identification of 
some period-bouncer candidates.  
Recent observations found that some of WZ Sge-type stars with 
type-B or type-E rebrightening are good candidates for the period 
bouncer and that the candidates share the following common 
observational features: (1) repeating rebrightenings or dips 
in brightness at the main superoutburst stage, (2) long-lasting 
stage A superhumps, (3) large decreases in the superhump period 
at the stage A to B transition in the objects with repeating 
rebrightenings, (4) small superhump amplitudes ($\lesssim$ 0.1 mag), 
(5) long delays of the appearance of ordinary superhumps, 
(6) slow fading rates at the plateau stage with ordinary superhumps 
in the superoutburst, and (7) large outburst amplitudes 
at the time of appearance of ordinary superhumps (e.g., 
\cite{nak14j0754j2304} and \cite{kim18a16dt}).  

In this paper, we report on our multi-wavelength photometric 
observations of the 2018 superoutburst of 
the famous WZ Sge-type star EG Cnc which was originally 
discovered by \citet{hur83egcnc}.  
This system underwent a superoutburst in 1996--1997 
\citep{pat98egcnc} and is supposed to have entered a normal 
outburst in 2009 \citep{tem09egcnc}.  
We started a photometric campaign of the 2018 superoutburst 
of this system soon after the detection of the luminosity 
increase by Patrick Schmeer.  
The mass ratio of EG Cnc was estimated to be 0.027--0.035 
by using the superhump period in its 1996--1997 superoutburst, 
which is the smallest among the identified WZ Sge stars 
\citep{pat98egcnc,pat11CVdistance}, 
and hence, EG Cnc is considered to be the best candidate 
for the period bouncer.  
However, the estimate was derived from the empirical relation 
between the excess of superhumps and the mass ratio 
\citep{pat05SH}.  
Also, EG Cnc showed 6 consecutive rebrightenings 
just after the main superoutburst in the past.  
Each rebrightening is similar to a normal outburst with 
an interval of about a week, and hence, this target is 
suitable for the detailed observations of rebrightenings.  
Our purpose in this paper is to more accurately estimate 
the mass ratio of EG Cnc to examine if it is a possible 
period-bouncer candidate or not and to constrain 
the origin of multiple rebrightenings.  
This paper is structured as follows.  We describe the methods of 
our observations and data analyses in Sec.~2 and present 
the results in Sec.~3.  
In Section 4, we discuss the results, and a summary and 
conclusions are given in Sec.~5.

\section{Observations and Analyses}

Time-resolved optical and near-infrared CCD photometry was 
carried out by the Variable Star Network (VSNET) collaboration 
team and the Optical and Infrared Synergistic Telescopes 
for Education and Research (Oister) at 21 sites.  
Table E1 shows the log of photometric observations.  
All of the observation times are converted to barycentric 
Julian date (BJD).  
Each observer performed differential photometry.  
We have applied zero-point corrections to each observer 
by adding constants before analyses.  
The constancy of the comparison star that each observer used 
was checked by nearby stars within the same images.  
The observational log is summarized in Table E1 in 
the supplementary information.  

The X-ray and ultraviolet ($UV$) observations were 
performed by Swift/X-ray Telescope (XRT) and 
Swift/UltraViolet and Optical Telescope (UVOT), 
respectively.  The ObsIDs are 00035446003--00035446042.  
The X-ray data are processed through xrtpipeline.  
The UV flux is obtained via the standard tool uvot2pha 
provided by the swift team.  
All of the observation times are converted to barycentric 
Julian date (BJD) as did in the optical and NIR data.  

We adopt the phase dispersion minimization (PDM) method \citep{PDM}
for period analyses.  Before applying PDM, we subtract 
the long-term trend of the light curve by locally weighted 
polynomial regression (LOWESS: \cite{LOWESS}).  
The 1$\sigma$ error of the best estimated period is determined 
by the same method as that described by \citet{fer89error} and 
\citet{Pdot2}.  
A variety of bootstraps was used for evaluating the robustness of 
the PDM result.  We make 100 samples which randomly contained 
50\% of observations, and perform PDM analyses for these 
samples.  
The result of the bootstraps is expressed as a form of 90\% 
confidence intervals in the resultant $\theta$ statistics.  
Also, we determine the times of maxima of 
ordinary superhumps in the same way as in \citet{Pdot}, 
when drawing the $O-C$ curve.

\section{Results}

\subsection{Overall light curves}

\begin{figure}[htb]
\begin{center}
\FigureFile(80mm, 50mm){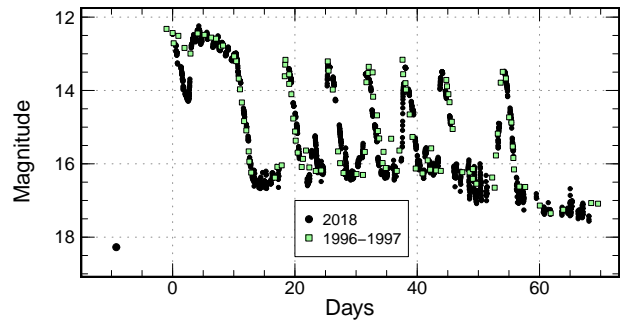}
\end{center}
\caption{Overall light curve of the 2018 superoutburst of EG Cnc (filled circles) compared with that in its 1996--1997 superoutburst (open rectangles).  Day 0 of the 2018 superoutburst is adjusted to BJD 2458396.9184, and that of the 1996--1997 superoutburst is adjusted to BJD 2450420.1529, respectively.  }
\label{overall}
\end{figure}

We show the overall light curve of the 2018 superoutburst of 
EG Cnc in Figure \ref{overall} with the $V$-band light curve 
of its 1996--1997 superoutburst for comparison, which is derived 
from Fig.~1 of \citet{pat98egcnc}.  
Although it is unclear when the 2018 superoutburst began, 
the luminosity increase would occur at least after BJD 2458387.7.  
After the increase, the luminosity suddenly dropped by $\sim$2 mag 
around BJD 2458399 at the plateau stage.  
After the dip, the plateau continued during a week.  
Multiple rebrightenings (type-B rebrightening) were confirmed 
soon after the main superoutburst.  
The rebrightening behavior including the timing of each outburst 
in the 2018 superoutburst seems to be completely the same as 
that in the previous superoutburst.  
However, the depth of the luminosity dip in the plateau stage 
is different between these two outbursts.  
The duration of the plateau stage might be different as well.

\subsection{Time evolution of superhumps}

\begin{figure}[htb]
\begin{center}
\FigureFile(80mm, 50mm){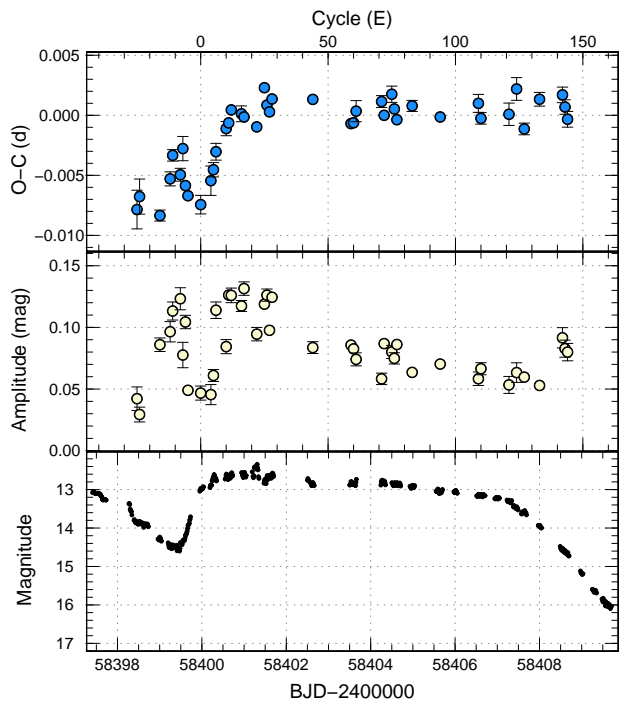}
\end{center}
\caption{Upper: $O - C$ curve of times of maxima of light modulations during BJD 2458398.4--2458408.7.  An ephemeris of BJD 2458399.97$+$0.06032015 E is used for drawing this figure.  Middle: amplitude of humps.  Lower: light curve. The horizontal axis in units of BJD and the cycle number is common to these three panels.
}
\label{oc}
\end{figure}

We have combined our $R$-band, $V$-band, and unfiltered data, and 
have made the $O - C$ curve of times of superhump maxima 
to \textcolor{black}{search for superhumps}.  
The $O - C$ curve before the rebrightening stage is given 
in the top panel of Figure \ref{oc} with the amplitude of 
superhumps and the light curve in its middle and bottom panels.  
The $O - C$ value is summarized in Table E2 in the supplementary 
information.  
Some points with large errors are removed from this figure.  
According to the past studies on some WZ Sge stars entering 
superoutbursts with a luminosity dip in the middle of the plateau, 
there is a possibility that EG Cnc exhibited early superhumps 
before the dip \citep{kim18a16dt}; however, we do not find them.  
It is unknown whether no detection of early superhumps 
is attributed to no observations or the fact that early superhumps 
did not develop.  

Small-amplitude light variations seem to have 
appeared on BJD 2458398 and the onset of superhumps would 
be \textcolor{black}{delayed with respect to the first 
appearance of the small-amplitude variations}.  
The linear trend starting from the cycle 0 in the $O - C$ curve 
and the increasing amplitude of humps suggest that superhumps 
began developing on the cycle 0 and that the interval 
during BJD 2458399.7--2458400.8 ($0 \leq E \leq 12$) is stage A.  
\textcolor{black}{
The $O-C$ curve bends around the cycle 12, which means that 
the superhump period suddenly decreased.  This indicates 
the onset of stage B superhumps.  
The period variation after this (see the texts in 
the next paragraph) and the decreasing amplitude of humps 
suggest that the interval during BJD 2458400.8--2458408.7 
($16 \leq E \leq 144$) is stage B.  }
However, the fluctuations in the $O - C$ curve before 
the cycle 0 is not consistent with the observational 
properties of superhumps and early superhumps, which 
have been well investigated (see e.g., \cite{Pdot} and 
\cite{kat15wzsge}).  

\begin{figure*}[htb]
\begin{minipage}{0.33\hsize}
\begin{center}
\FigureFile(55mm, 50mm){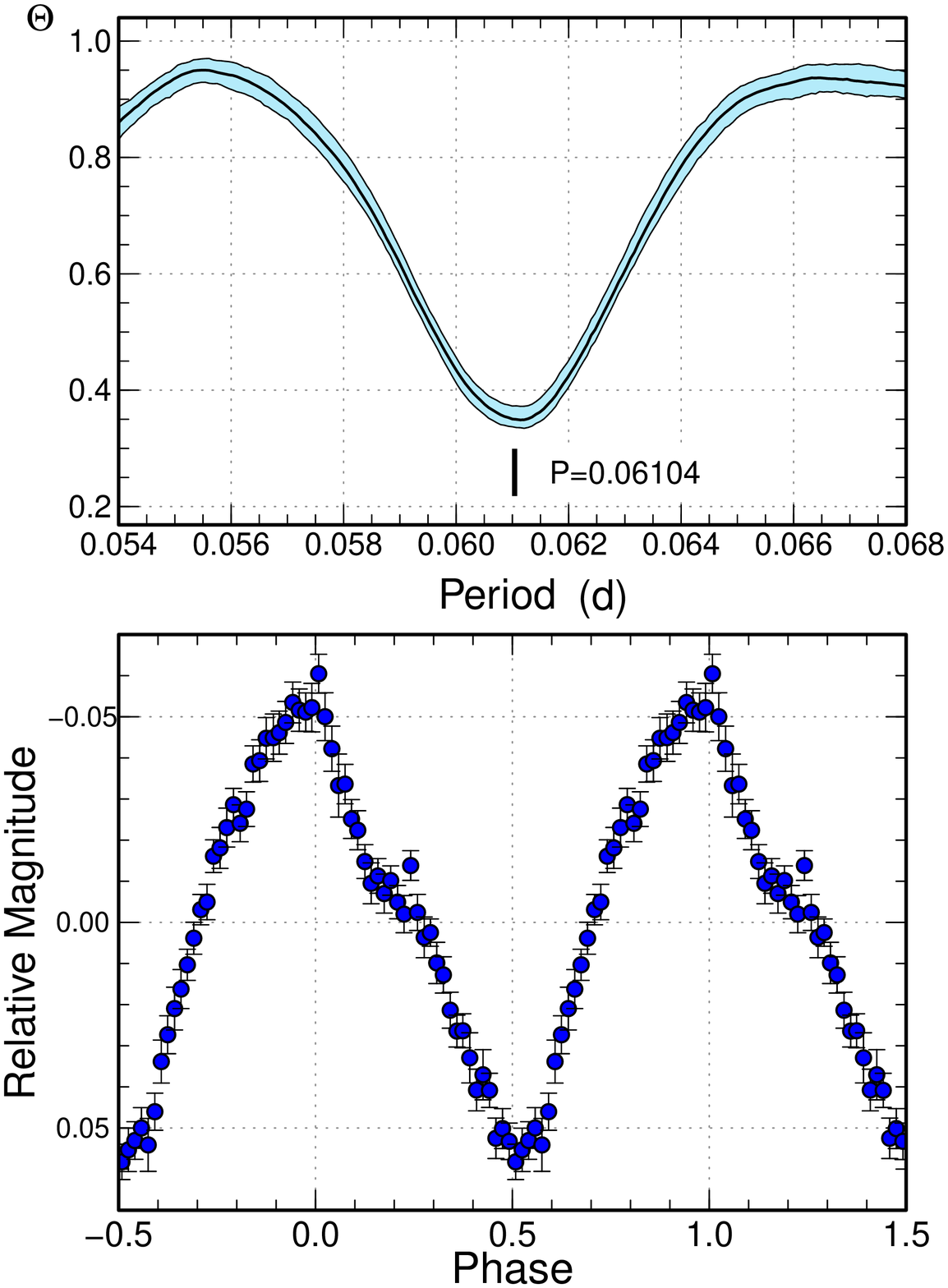}
\end{center}
\end{minipage}
\begin{minipage}{0.33\hsize}
\begin{center}
\FigureFile(55mm, 50mm){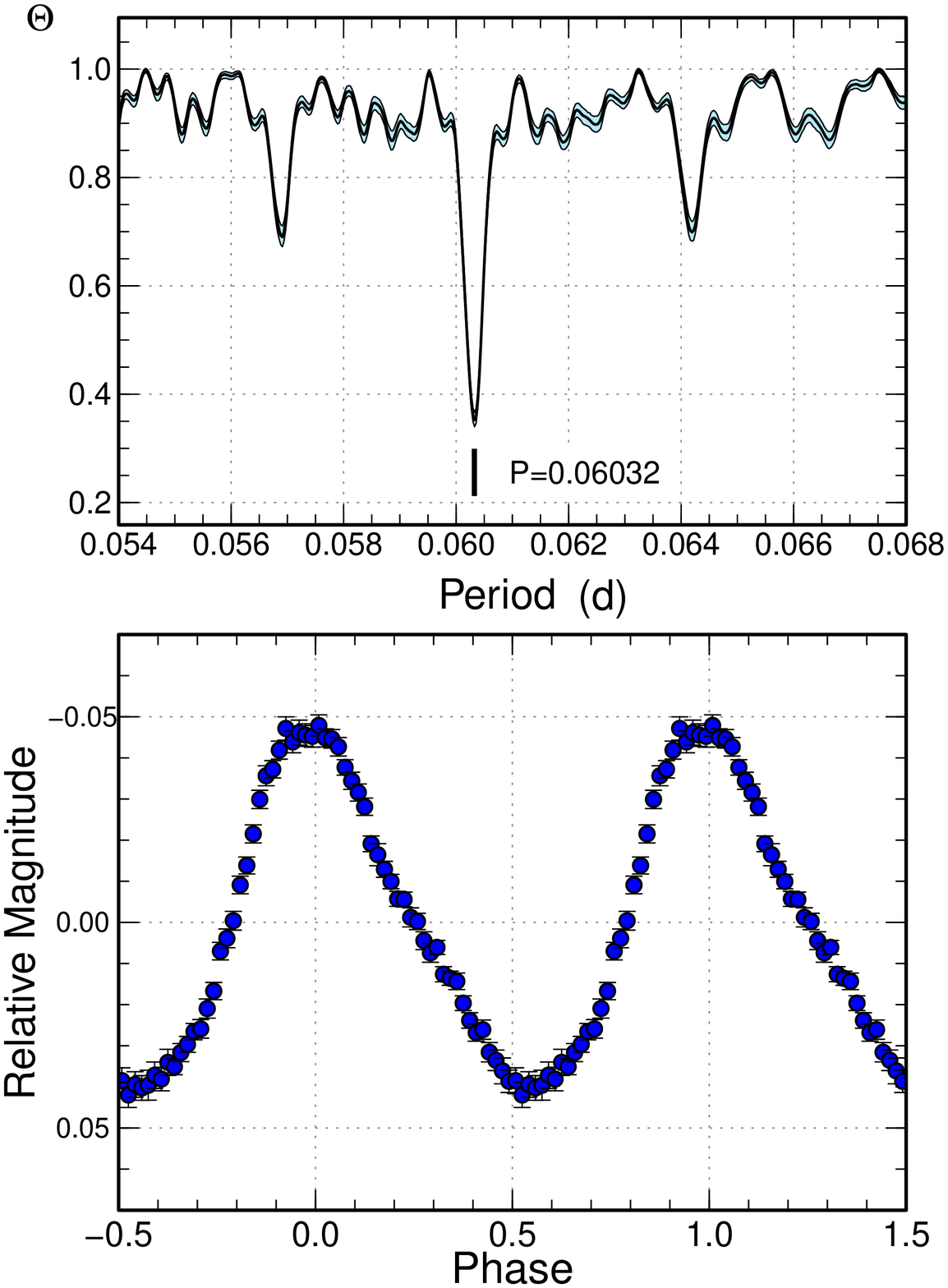}
\end{center}
\end{minipage}
\begin{minipage}{0.33\hsize}
\begin{center}
\FigureFile(55mm, 50mm){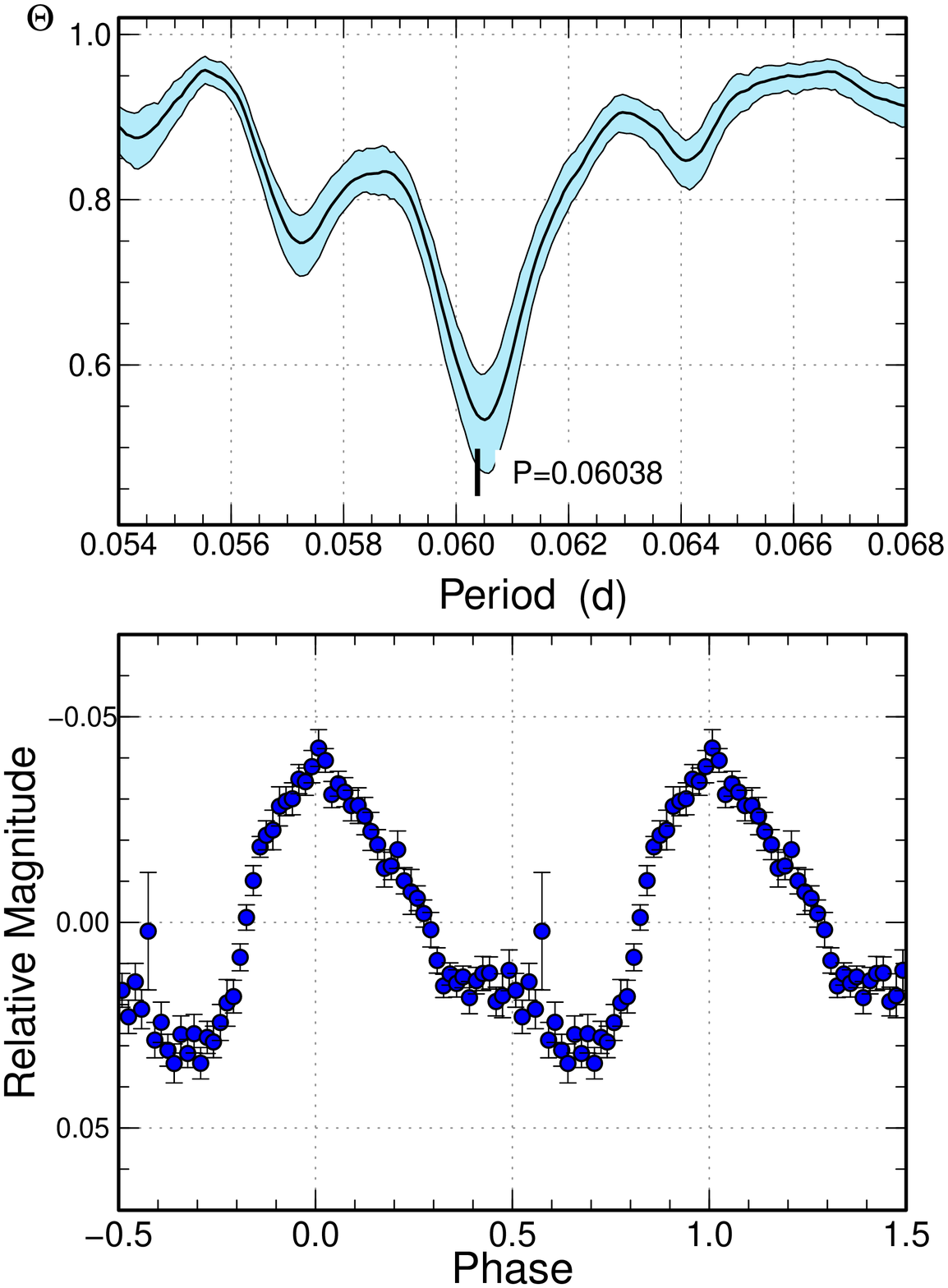}
\end{center}
\end{minipage}
\\
\begin{minipage}{0.325\hsize}
\begin{center}
\FigureFile(55mm, 50mm){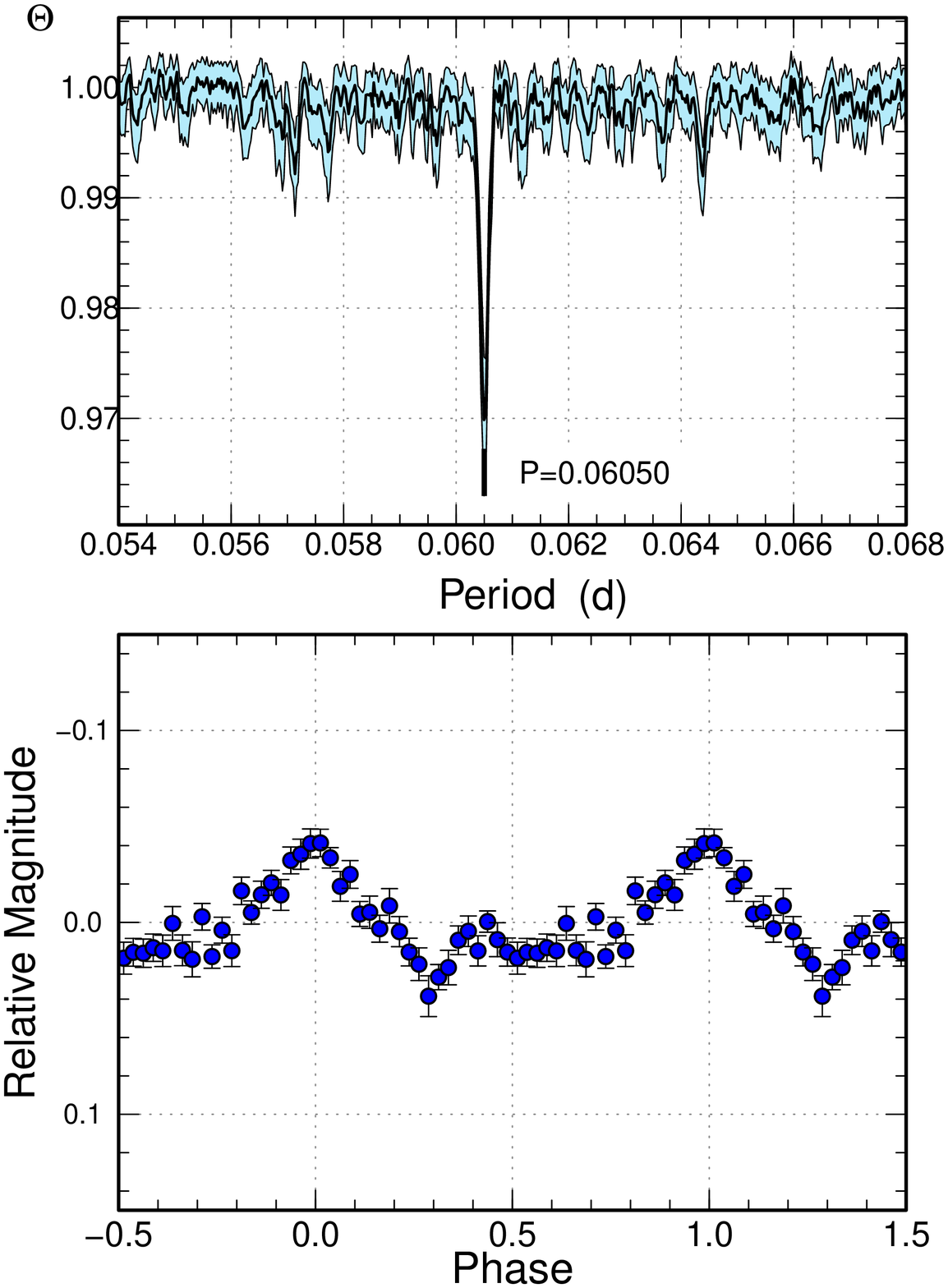}
\end{center}
\end{minipage}
\begin{minipage}{0.325\hsize}
\begin{center}
\FigureFile(55mm, 50mm){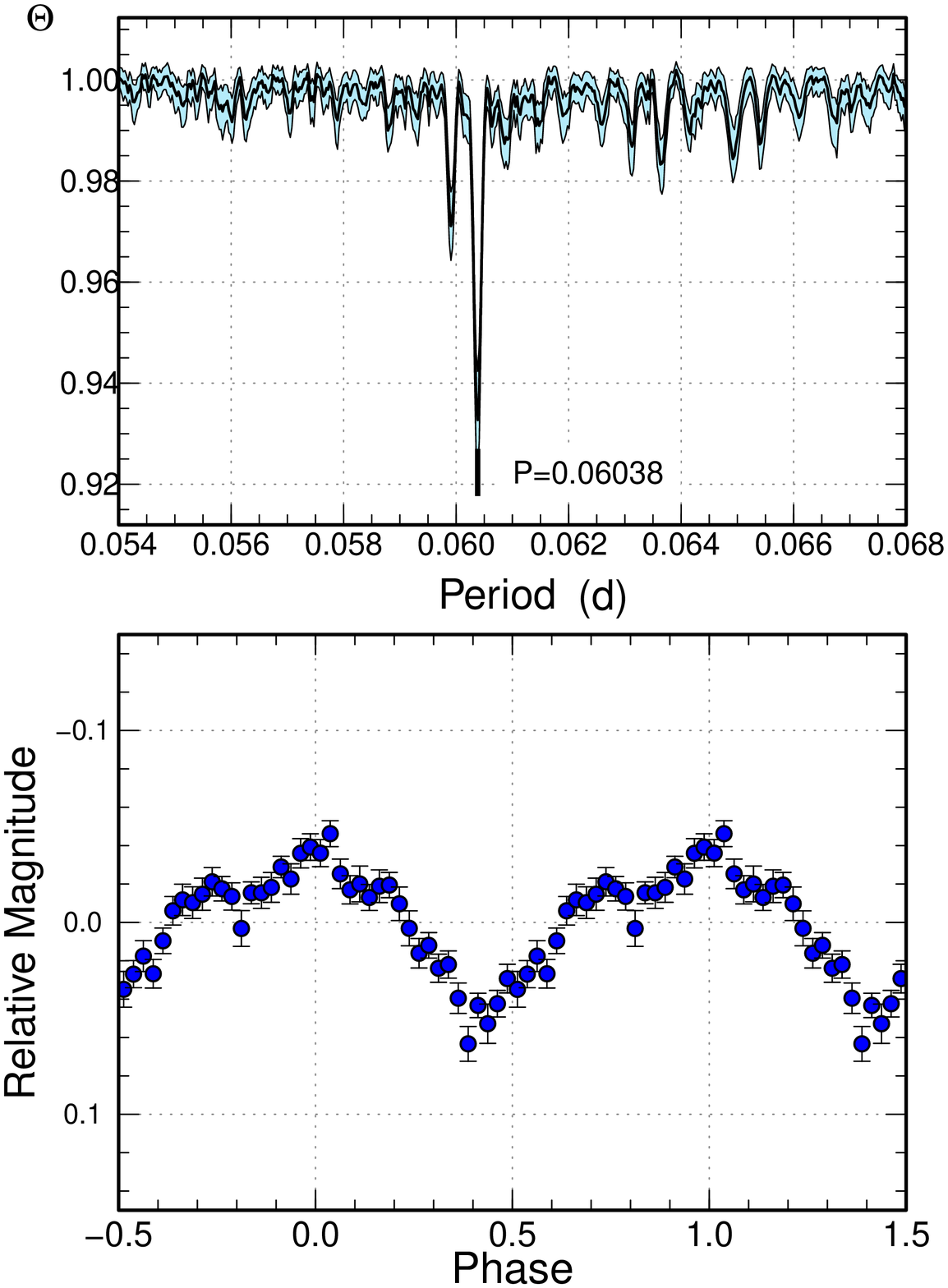}
\end{center}
\end{minipage}
\begin{minipage}{0.325\hsize}
\begin{center}
\FigureFile(55mm, 50mm){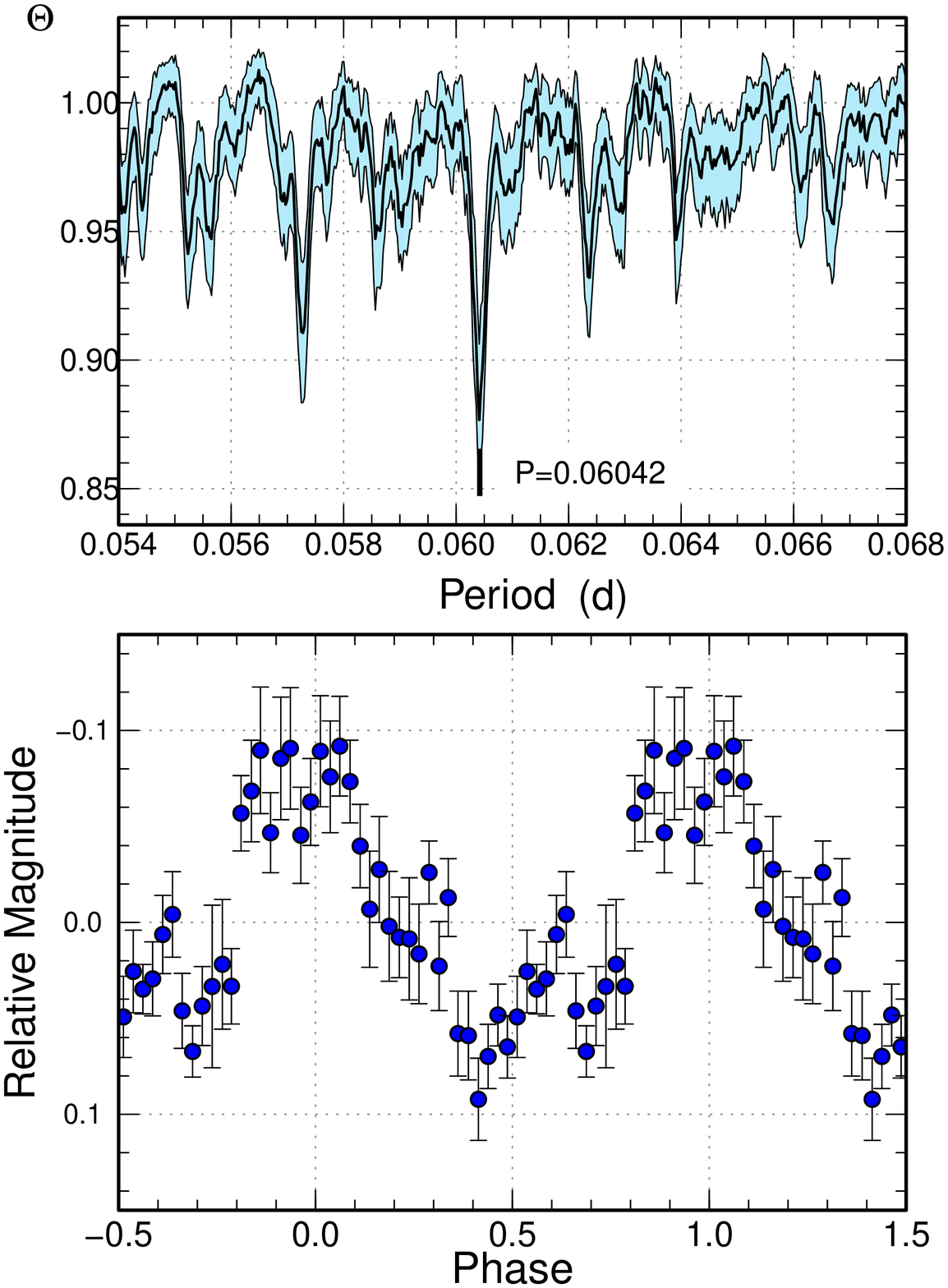}
\end{center}
\end{minipage}
\vspace{5mm}
\caption{Stage A superhumps, stage B superhumps, light modulations before stage A, stage La superhumps, stage Lb superhumps, and stage Lc superhumps in the 2018 superoutburst of EG Cnc.  The upper panels are $\Theta$-diagrams of our PDM analysis and the lower panels are phase-averaged profiles, respectively.  Upper left: stage A superhumps (BJD 2458399.7--2458400.8).  Upper middle: stage B superhumps (BJD 2458400.8--2458408.7).  Upper right: light modulations before stage A (BJD 2458398.4--2458399.7).  Lower left: stage La superhumps (BJD 2458412.9--2458431.7).  Lower middle: stage Lb superhumps (BJD 2458435.5--2458453.7).  Lower right: stage Lc superhumps (BJD 2458457.5--2458471.7).  }
\label{pdm}
\end{figure*}

The PDM results and the phase-averaged profiles of 
\textcolor{black}{stage A superhumps, stage B superhumps, and 
light modulations before stage A} 
are displayed in \textcolor{black}{the upper left, 
upper middle, and upper right panels} of 
Figure \ref{pdm}.  These modulations are clearly single-peaked.  
The stage A superhump period ($P_{\rm shA}$) estimated by PDM is 
0.06103(2)~d (see \textcolor{black}{the upper left panel of} 
Figure \ref{pdm}), and the averaged period at stage B 
($P_{\rm shB}$) estimated by PDM is 0.060324(3)~d, respectively 
(see \textcolor{black}{the upper middle panel of} Figure \ref{pdm}).  
The derivative of the superhump period during stage B 
($P_{\rm dot} (\equiv \dot{P}_{\rm sh}/P_{\rm sh})$) is 
$-1.5 (4.0) \times 10^{-6}~{\rm s}~{\rm s}^{-1}$.  
The averaged superhump amplitude during the plateau stage is 0.08~mag.  
On the other hand, the humps before the onset of stage A superhumps 
have a $\sim$0.0604-d period, which is shorter than the stage A 
superhump period and longer than the orbital period 
0.05997(9)~d \citep{pat98egcnc} 
(see \textcolor{black}{the upper right panel of} Figure \ref{pdm}).  
They do not seem to be either superhumps or early superhumps 
from their periodicity.  

\begin{figure}[htb]
\begin{center}
\FigureFile(80mm, 50mm){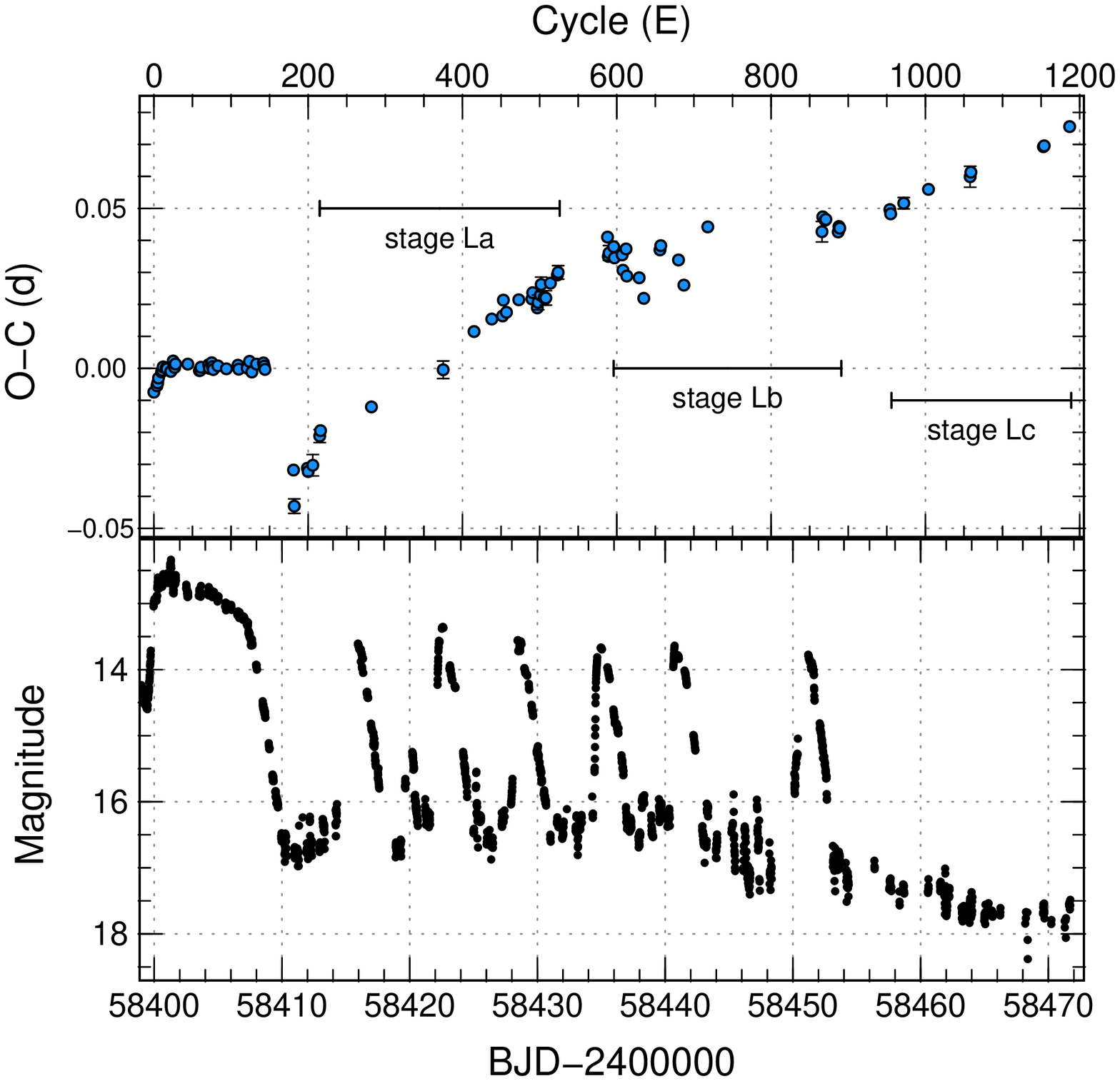}
\end{center}
\caption{$O - C$ curve of the times of superhump maxima and light curves during BJD 2458399.7--2458471.7.}
\label{oc-all}
\end{figure}

The superhumps seem to have continued for a long time 
after the main superoutburst.  
Long-lasting superhumps have been observed previously 
in WZ Sge stars including EG Cnc itself in the past 
\citep{pat98egcnc,pat02wzsge,kat08wzsgelateSH,neu17j1222,tam20j2104}.  
Figure \ref{oc-all} shows the $O-C$ curve of times of 
superhump maxima extending to the cycle $\sim$1200.  
Long-lasting late-stage superhumps seem to have begun 
at the cycle 215 in the 2018 superoutburst of EG Cnc.  
The superhumps in the cycles 181--200 may be of the end of 
stage C.  As seen in some other WZ Sge-type superoutbursts, 
$\sim$0.5 phase shift in superhumps seem to have 
occurred when this system entered the post-stage of 
the main superoutburst (see also \cite{kat15wzsge}) 
and this phenomenon was also confirmed in the 1996--1997 
superoutburst of this object \citep{kat04egcnc}.  
This evolution of late-stage superhumps in the $O-C$ curve 
would be similar to that in the 1996--1997 superoutburst 
in EG Cnc (see also \cite{pat98egcnc}).  
Both of the $O-C$ curves seem to smoothly bend around 
the cycle 600.  
We can see that the superhump period fluctuated in 
the middle of the late-stage and that the period became 
constant again after that.  
We thus divide the late-stage superhumps into the three 
intervals according to their period variations: 
during BJD 2458412.9--2458431.7 ($215 \leq E \leq 526$) 
as stage La, during BJD 2458435.5--2458453.7 ($590 \leq E \leq 891$) 
as stage Lb, and during BJD 2458457.5--2458471.7 ($956 \leq E 
\leq 1189$) as stage Lc.  
It is difficult to distinguish stage Lb and stage Lc, 
and we determine the time interval of stage Lc 
to minimize $P_{\rm dot}$.  
We give the PDM results and the phase-averaged profiles of 
stage La, stage Lb, and stage Lc superhumps in 
\textcolor{black}{the lower left, lower middle, and 
lower right panels of} Figure \ref{pdm}.  
The superhump periods during stage La and stage Lc are 
0.06050(1)~d and 0.06042(2)~d, respectively.  
The averaged superhump period in stage Lb is 0.060383(7)~d, 
and the derivative of the superhump period during stage B 
($P_{\rm dot}$) is $4.8 (5.6) \times 10^{-5}~{\rm s}~{\rm s}^{-1}$.

\subsection{Multi-wavelength behavior and color variations}

\begin{figure}[htb]
\begin{center}
\FigureFile(80mm, 50mm){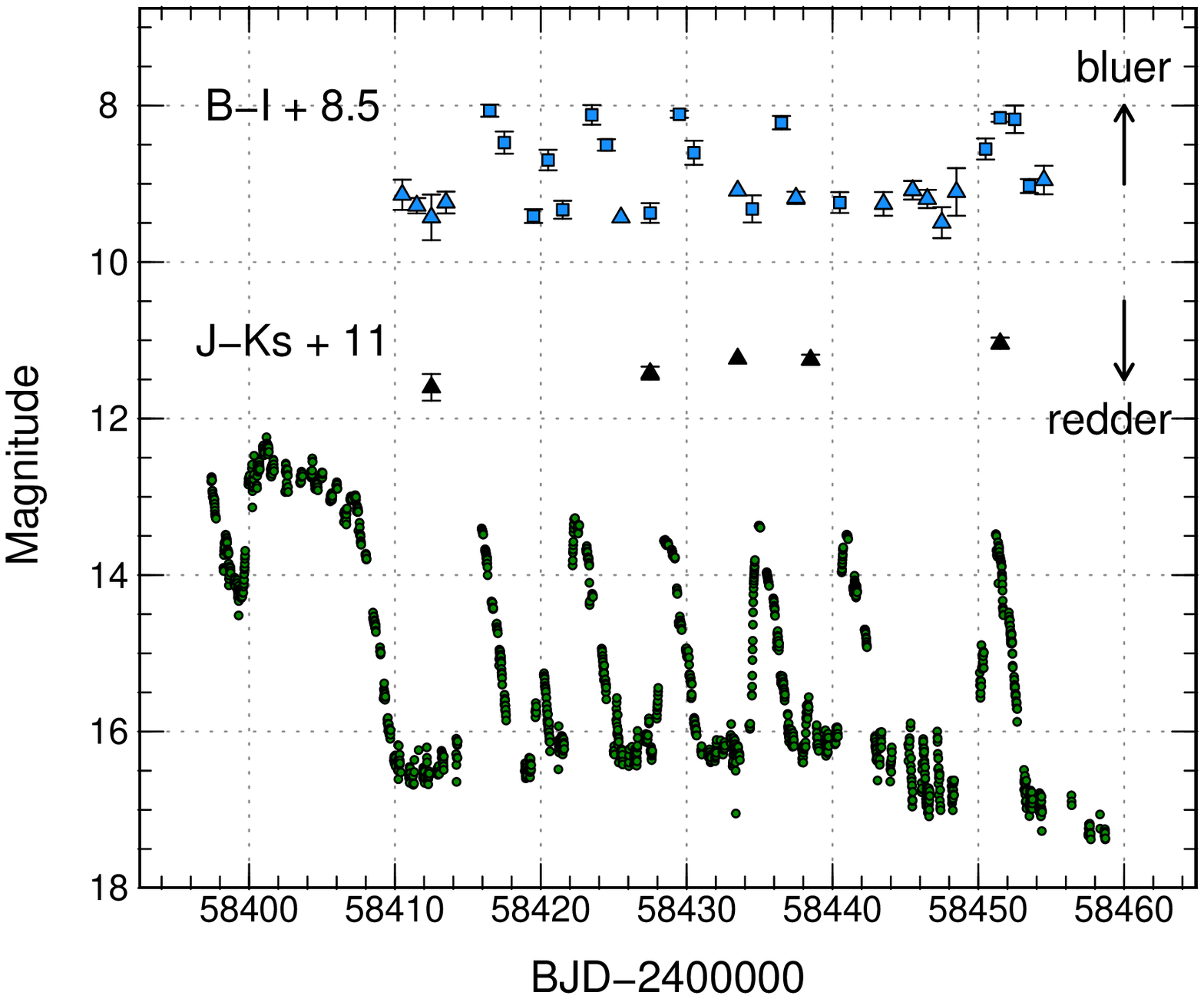}
\end{center}
\caption{Overall $V$-band light curve of the 2018 superoutburst of EG Cnc, and variations of $B-I$ and $J-K_{\rm S}$ colors.  The triangles and rectangles represent the color in quiescence between rebrightenings and that in each rebrightening, respectively.  }
\label{overall-color}
\end{figure}

We have explored the daily variation of optical and NIR colors 
during the rebrightening stage of the 2018 superoutburst in EG Cnc.  
\textcolor{black}{
In Figure \ref{overall-color}, we show daily averages of 
the colors and the $V$-band light curve.  }
The $B-V$ and $J-K_{\rm S}$ colors are extracted from 
the data obtained by KU1 and HHO, respectively.  
During quiescence between rebrightenings, 
the averaged $B-I$ color was 0.8, which is redder than 
that in quiescence in typical DNe \citep{bai80DNcolor}.  
This phenomenon is consistent with the unusually red $V-I$ color 
in the rebrightening phase of the previous superoutburst of 
this object \citep{pat98egcnc}.  
Also, the $J-K_{\rm S}$ color became bluer.  
On the other hand, the $B-I$ color became bluer 
during each rebrightening.  

Let us consider the continuum emission from a WZ Sge star.  
In WZ Sge stars, the radiation from the accretion disk is 
dominant over UV, optical, and NIR wavelengths, 
since the white dwarf and the secondary star are tiny and 
since the boundary layer emits X-rays.  
If the disk is optically thick and geometrically thin, 
the emission from the disk surface is approximately 
blackbody radiation.  
Generally, the temperature of the inner part is high and 
that of the outer part is low in the accretion disk 
since the gravitational potential of the white dwarf 
becomes deeper inwards in the disk.  
Simply, each annulus in the disk has each different 
temperature and multi-temperature blackbody emission 
synthesizes the radiation from the disk.  
In fact, the observed continuum spectra of the disk 
around the outburst maximum agree well with multi-color 
blackbody emission expected from the standard-disk model 
\citep{hor85zcha,sha73BHbinary}.  
It is considered that UV photons come mainly from 
the inner part of the disk.  
A Rayleigh-Jeans slope of the multi-color blackbody 
typically exists around optical and NIR wavelengths 
in the spectrum, which depends on the disk radius and 
the temperature of the outermost part of the disk.  
The $B-I$ color is sensitive to the temperature distribution 
of the outer disk and the $J-K_{\rm S}$ color is affected 
by the activity at the lower-temperature outermost 
region of the disk and the time variation of the disk radius.  
The $B-I$ color of 0.8 in our observations
corresponds to the black-body radiation of $\sim$7000~K.  
Also, the $J-K_{\rm S}$ color of 0.2--0.6 corresponds to 
the black-body radiation of $\sim$4000--7000~K.  

According to the picture described above, the red $B-I$ 
and $J-K_{\rm S}$ colors at the beginning of 
the rebrightening stage suggests the presence of 
the cool component at the outermost disk.  
If the low-temperature gas at the outermost disk is depleted 
with accretion during rebrightenings, i.e., the disk shrinks 
with time, the $J-K_{\rm S}$ color in the quiescent state 
between rebrightenings should become bluer.  
The bluer $J-K_{\rm S}$ color in our observations agrees 
with this idea.  
On the other hand, the bluer $B-I$ color during each 
rebrightening implies that the temperature of the outer disk 
increased at that time, because this region goes up to 
the outburst state.  
The $B-I$ color of 0.0 corresponds to the black-body radiation 
of $\sim$18000~K, which is higher than the minimum temperature 
of the outburst state (see e.g., \cite{min83DNDI}).  
For the solid confirmation, we need to analyze the spectral 
energy distribution of the disk, which is beyond the scope 
of this paper.  

\begin{figure*}[htb]
\begin{minipage}{0.49\hsize}
\begin{center}
\FigureFile(70mm, 50mm){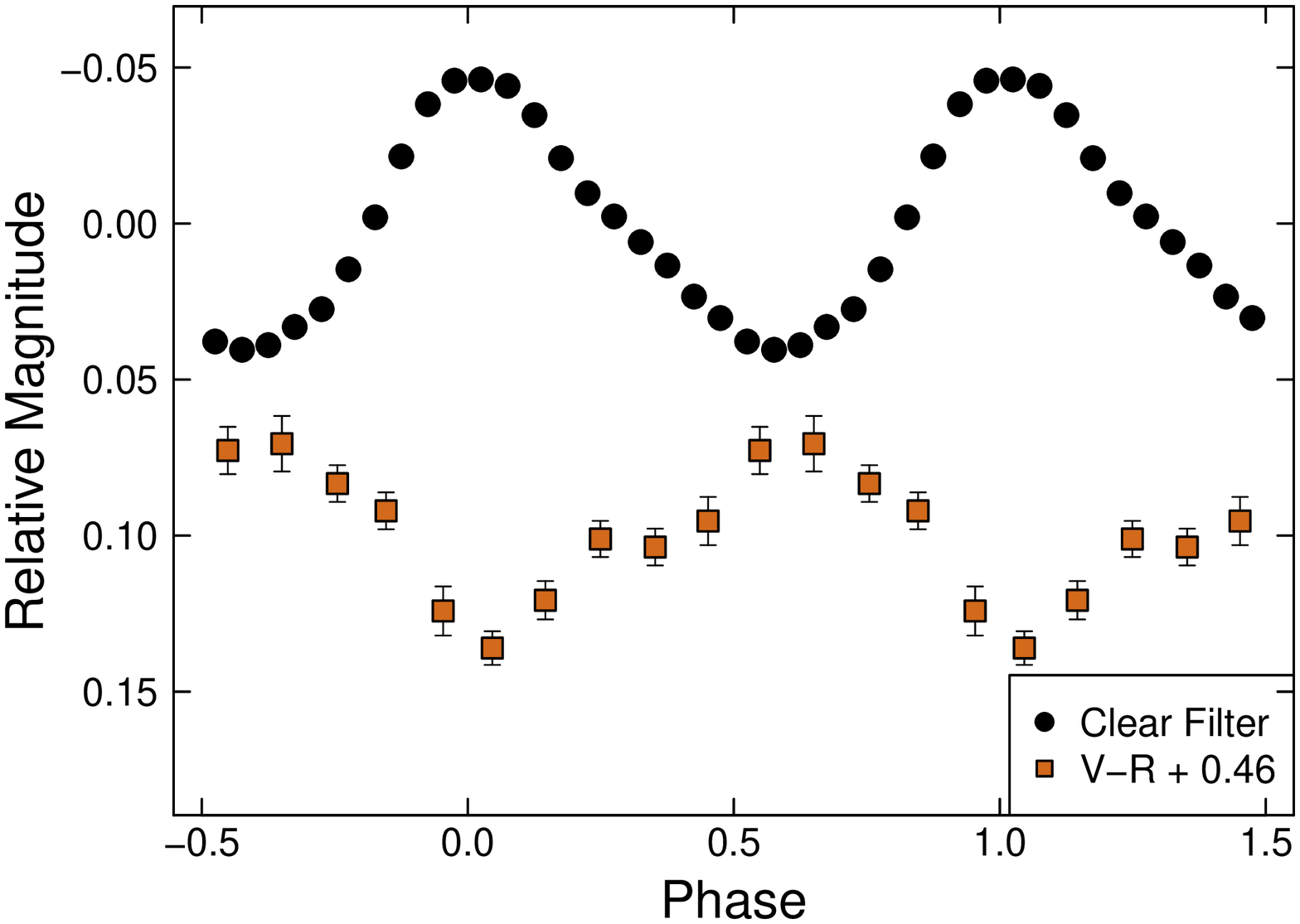}
\end{center}
\end{minipage}
\begin{minipage}{0.49\hsize}
\begin{center}
\FigureFile(70mm, 50mm){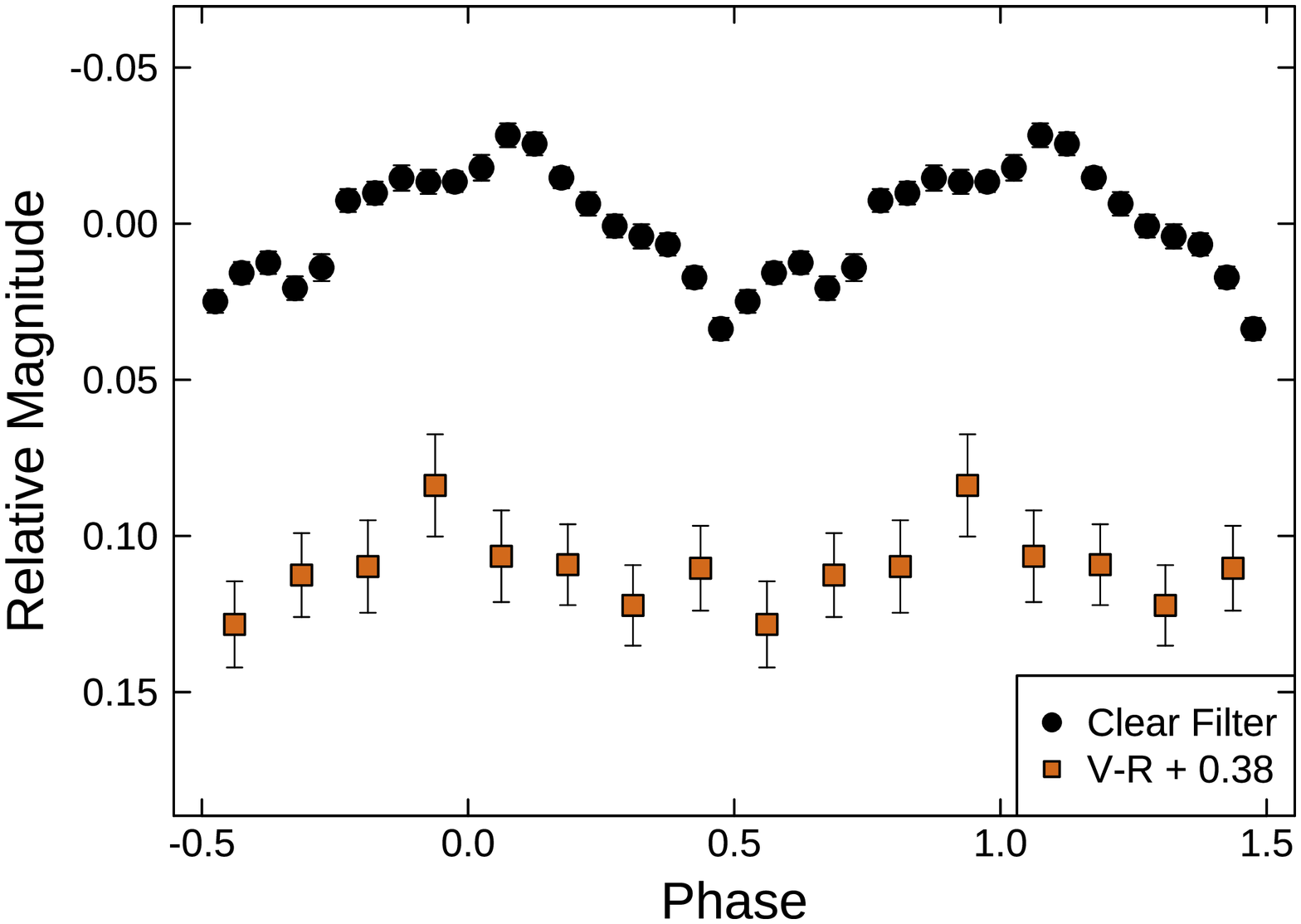}
\end{center}
\end{minipage}
\vspace{8mm}
\caption{Phase-averaged profile of superhumps and $V-R$ color.  Left: stage B superhumps (BJD 2458400.8--2458408.7).  Right: late-stage superhumps in quiescence between rebrightenings (BJD 2458412.9--2458471.7).}
\label{SHcolor}
\end{figure*}

We also have investigated the color variation of 
the phase-averaged profile of stage B and late-stage 
superhumps.  
The phase-averaged profiles and the $V-R$ colors are 
exhibited in Figure \ref{SHcolor}.  
The color became redder at the maximum of superhumps 
during stage B, while it became bluer after the main 
superoutburst.  

\begin{figure}[htb]
\begin{center}
\FigureFile(80mm, 50mm){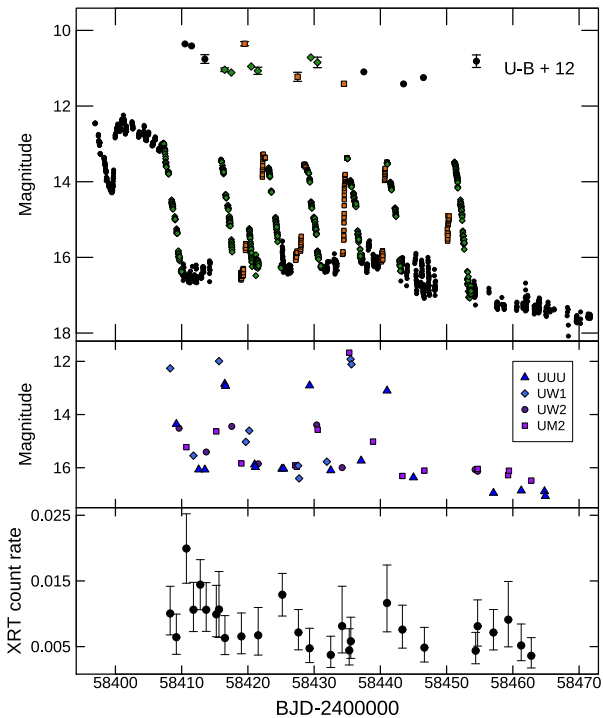}
\end{center}
\caption{Optical, $UV$, and X-ray light curves of the 2018 superoutburst of EG Cnc.  In the top panel, we denote the fading and the rising parts as diamonds and rectangles, respectively.  Also, we give the $U-B$ color in the top panel, which is denoted as different symbols corresponding to those of the light curve.  }
\label{multi}
\end{figure}

\begin{figure}[htb]
\begin{center}
\FigureFile(80mm, 50mm){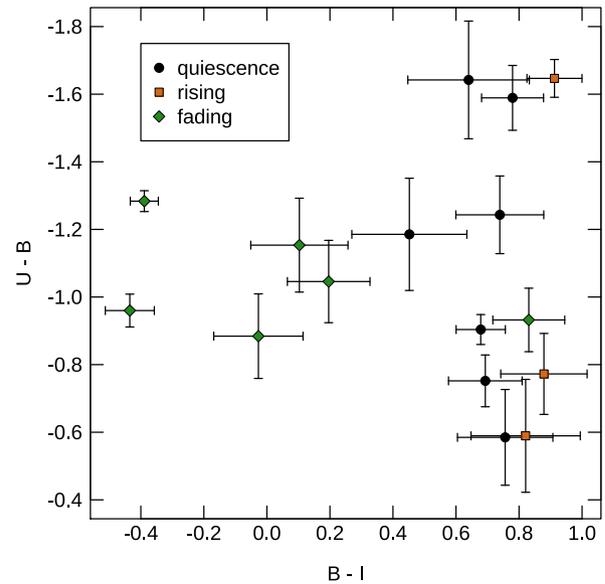}
\end{center}
\caption{Two-color plane composed of the $U-B$ and $B-I$ colors during the rebrightening phase.  The circles, rectangles, and diamonds represent the data during quiescence between rebrightenings, the rising part, and the fading part, respectively.  }
\label{ub-bi}
\end{figure}

Moreover, we have obtained multi-wavelength light curves during 
rebrightening by using the {\it Swift} satellite.  
The optical, $UV$, and X-ray light curves are 
exhibited in Figure \ref{multi}.  
This system was faint at X-ray wavelengths during 
the rebrightening stage, which suggests that the boundary 
layer between the inner disk rim and the white dwarf 
became optically thick during multiple rebrightenings 
because of the high accretion rate to the white dwarf 
(e.g., \cite{pri79dnexray,pat85CVXrayemission1}).  
On the other hand, the $UV$ light curve is sensitive to 
multiple rebrightenings and seems to well trace the optical one.  
The daily $U-B$ color is also displayed in the top panel of 
Figure \ref{multi}.  
Although the $UV$ observations were performed by 4 different bands, 
we here treat them as the same $U$ band for simplicity.  
The $U-B$ color became redder at least 
during quiescence between rebrightenings.  
As described in the second paragraph of 
this section, the inner part of the accretion disk is 
generally hotter than its outer part and $UV$ photons 
are regarded to be released from the innermost region of the disk.  
The $U-B$ color is thus considered to be sensitive to the activity 
in the inner disk as expected by numerical simulations 
\citep{can87DNspecevolution}.  
The redder $U-B$ color during the rebrightening phase may 
represent that the inner region of the disk becomes cooler.  
The $U-B$ color seems to be at a similar level in comparison 
with the observed $U-B$ colors in other dwarf novae 
\citep{bai80DNcolor,ech83DNphotometry,shu18rzlmi}.  

We also show a diagram of the daily $U-B$ and $B-I$ colors 
during the rebrightening phase in Figure \ref{ub-bi}.  
\textcolor{black}{
We separate the colors into three intervals: the quiescence 
between rebrightenings, the rising part, and the fading part.  
The $U-B$ color during the rising part and quiescence 
has a large dispersion, because that color became redder 
during the rebrightening phase as described above.  }
Despite the large error bars, we see the trend that 
the $B-I$ colors are reddest at the rising part, and are 
bluest at the fading part.  
The $B-I$ color was slightly redder at the rising part 
of each rebrightening than that in quiescence and would 
represent the change in the temperature distribution of 
the outer disk and the disk radius.  
In the disk-instability model, the disk rapidly expands 
at the onset of outbursts and the cool outer region 
temporarily expands before the outer disk edge completely 
goes up to the hot state \citep{ich92diskradius} 
and the color should become redder 
only at the very early stage of the outbursts.  
Our observations are consistent with this picture.  
On the other hand, the $B-I$ color around the light maximum 
and/or at the fading part became much bluer, although there is 
one exception observed at the small brightening between 
the first and the second rebrightenings.  
This behavior agrees with the interpretation that 
the outer disk became hotter during each rebrightening, 
as described in the third paragraph of this subsection.  
The small rebrighetening between the first and the second 
rebrightenings is likely to have been limited in 
the narrow region of the disk at that time.  
Then the temperature of the outer disk does not rise very much, 
which keeps the $B-I$ color red.

\section{Discussion}

\subsection{Is EG Cnc one of the best candidates for the period bouncer ?}

We firstly estimate the mass ratio of EG Cnc by 
using the stage A superhump period that we have derived 
(see Sec.~3.2) and the orbital period (0.05997(9)~d) 
reported by \citet{pat98egcnc}, complying with the method 
proposed by \citet{kat13qfromstageA}.  
The superhump period is basically determined by the dynamical 
precession of the eccentric disk and the pressure effect 
decreasing the precession rate \citep{lub92SH,hir93SHperiod}, 
and hence, the apsidal precession rate of the eccentric disk, 
$\omega_{\rm pr}$, is approximately formulated as
\begin{equation}
\omega_{\rm pr} = \omega_{\rm dyn} + \omega_{\rm press}, 
\label{sh-freq}
\end{equation}
where $\omega_{\rm dyn}$ and $\omega_{\rm press}$ are 
the dynamical precession rate and the precession rate 
by the pressure effect, respectively.  
Stage A superhumps are the developing superhumps and are 
believed to represent the dynamical precession at 
the 3:1 resonance radius.  
According to \citet{hir90SHexcess}, the dynamical precession 
rate, is expressed as: 
\begin{equation}
\omega_{\rm dyn} / \omega_{\rm orb} = Q(q)R(r), 
\label{omega-dyn}
\end{equation}
where $Q(q)$ and $R(r)$ are the functions of the mass ratio 
and that of a given radius in the accretion disk, respectively 
(see equations (1) and (2) in \citet{kat13qfromstageA} for 
details), and $r$ is expressed as $(3^2(1+q))^{-1/3}a$, which is 
the 3:1 resonance radius in the case of stage A superhumps.  
Here $a$ is the binary separation.  
By using equation (\ref{omega-dyn}), the mass ratio of 
EG Cnc is estimated to be 0.045(5) which is consistent with 
the $P_{\rm dot}$ value derived in Sec.~3.2 and 
the type-B rebrightening (see also Fig.~17 in 
\cite{kat15wzsge}).  

However, the period of stage A superhumps might be diluted 
by other light modulations before stage A (see Sec.~3.2).  
For confirmation, we have constrained the mass ratio by using 
late-stage superhumps.  
The constant period of late-stage superhumps observed 
a long time after the main superoutburst would represent 
the dynamical precession at the outer rim of the disk as 
suggested by \citet{kat13qfromstageA}.  
If we know the mass ratio, the disk radius can be estimated 
from equation (\ref{omega-dyn}).  
We thus can constrain the mass ratio by using the period of 
stage Lc superhumps, if we determine the range of the disk radius.  
According to \citet{kat13qfromstageA}, the disk radius after 
the rebrightening phase in WZ Sge stars ranges between 
the Lubow-Shu radius \citep{lub75AD} and the 3:1 resonance radius.  
The derived limit of the mass ratio of EG Cnc is 0.019--0.057 
with 95\% confidence intervals.  
We take the median value as 0.048 since the disk radius long 
after the superoutburst accompanied by type-B rebrightening is 
empirically 0.30--0.32$a$ \citep{kat13qfromstageA}, which 
is consistent with the estimated mass ratio in the previous 
paragraph within 1$\sigma$ errors.  

\begin{figure}[htb]
\begin{center}
\FigureFile(80mm, 100mm){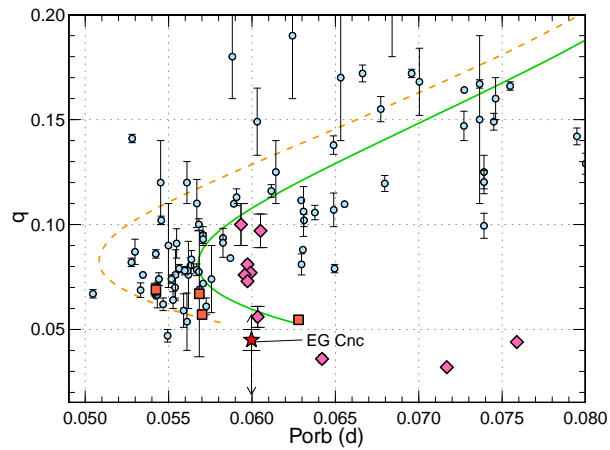}
\end{center}
\caption{$q-P_{\rm orb}$ relation of the candidates for the period bouncer and ordinary WZ Sge-type DNe.  The star, diamonds, rectangles and circles represent EG Cnc, other candidates for the period bouncer among the identified WZ Sge-type DNe, the candidates for the period bouncer among eclipsing CVs, and ordinary WZ Sge-type DNe.  The possible range of the mass ratio of EG Cnc is represented as the arrow underlying the star.  The dash and solid lines represent an evolutionary track of the standard scenario and that of the modified evolutionary theory, respectively, which are derived from \citet{kni11CVdonor}.}
\label{massratio}
\end{figure}

Our results thus allow a larger mass ratio for EG Cnc 
in comparison with the estimates in \citet{pat98egcnc} and 
\citet{pat11CVdistance}.  
The mass ratio of this object would be underestimated 
in the past since the period of stage B superhumps 
was used for the estimation, as pointed out by 
\citet{nak13j2112j2037}.  
The evolutionary stage of EG Cnc is exhibited in 
the $q - P_{\rm orb}$ plane as Figure \ref{massratio} 
based on our estimates of the mass ratio.  
This object still seems one of the good candidates for the period 
bouncer, since the mass ratio and the orbital period are similar 
to those of other period-bouncer candidates (the diamonds in 
Figure \ref{massratio}), even if we consider the wide range of 
the possible mass ratio.  

Also, we have found some characteristic observational 
features which period-bouncer candidates share 
in the 2018 superoutburst of EG Cnc.  
The mean amplitude of superhumps in the main superoutburst 
was smaller than that of ordinary WZ Sge-type 
stars (more than 0.1~mag) (see also Sec.~3.2).  
The fading rate of the middle of the plateau stage 
in which stage B superhumps were observed was small, 
0.098~mag~d$^{-1}$, which indicates a slow decline 
(see also \cite{kim18a16dt}).  
These properties are considered to originate from 
weak tidal dissipations due to extremely low-mass 
secondary stars.  This idea was already described in 
detail by \citet{kim18a16dt}.  

However, the recurrence time of superoutbursts and 
the duration of stage A superhumps in EG Cnc are shorter than 
those of the other candidates (see e.g., 
\cite{nak14j0754j2304,kim18a16dt}).  
Although the duration of stage A superhumps was possibly longer 
than that we have presented in Sec.~3.2, it is still less than 
a few days even if we consider the zero point of the superhump 
amplitude expected from the time evolution of the amplitude 
given in the middle panel of Figure \ref{oc}.  
Since the long duration of stage A superhumps is considered 
to originate from the very small tidal effect by the low-mass star, 
and since the long recurrence time of superoutbursts would be 
due to the very low mass-transfer rate from the low-mass secondary, 
the short duration of stage A superhumps and the short recurrence 
time of superoutbursts in EG Cnc seem to be inconsistent with 
its low binary mass ratio.  
Although it is unclear what this inconsistency comes from, 
as for the short recurrence time, the sustained eccentricity 
in the disk expected from the long-lasting superhumps 
in this object might contribute because the tidal dissipation 
of the disk by the secondary star increases the viscosity 
in the disk \citep{ich94tidal}.  
The temporal increase of the viscosity in the elliptical disk 
may be long-lasting even after superoutbursts, although 
the viscosity would gradually become lower with time.

\subsection{Luminosity dip in the plateau stage}

\citet{kim16a15jd} proposed that the slow development of 
the 3:1 resonance tidal instability cannot keep the efficient 
removal of the angular momentum from the outer disk after 
the fading of the 2:1 resonance due to weak tidal torques by 
the small secondary star in extremely low-$q$ objects, 
which makes a luminosity dip in the middle of the plateau stage 
in the main superoutburst.  
However, the dip in the 2018 superoutburst in EG Cnc may 
be caused by a different reason, since the disk mass accumulated 
before that superoutburst seems to be small.  
EG Cnc entered one normal outburst between the 1996--1997 
superoutburst and the 2018 superoutburst and some of the disk mass 
seems to have been depleted at that time.  
Although the disk is believed to expand far from the 2:1 resonance 
radius at the onset of superoutbursts in extremely low-$q$ objects, 
period-bouncer candidates, there may be a possibility that the disk 
radius did not exceed beyond the 2:1 resonance in the 2018 
superoutburst of EG Cnc because of the small amount of 
mass stored in the disk.  

We thus consider that the brightening before the dip in 
the 2018 outburst of EG Cnc is probably a precursor, 
i.e., a normal outburst triggering a superoutburst, which is 
usually observed before the superoutbursts in SU UMa-type 
stars (see e.g., \cite{osa13v1504cygKepler}), as confirmed 
in the 2015 superoutburst of AL Com, another WZ Sge star, 
which was triggered after only 1.5 yr quiescence \citep{kim16alcom}.

\subsection{Implications to the mechanism of rebrightenings}

The light curve in the rebrightening stage of the 2018 
superoutburst in EG Cnc was completely the same as that 
of its previous superoutburst in 1996--1997 despite 
the difference in the light curve of the main superoutburst 
(see Figure \ref{overall}).  
This is the second example following 
the 2015 superoutburst in AL Com \citep{kim16alcom}.  
In the case of AL Com, the duration of the type-A (long duration) 
rebrightening was a little shorter in its 2015 superoutburst 
than that in its 2013 superoutburst (see also \cite{Pdot6}).  
However, not only the rebrightening type but also the timing of 
each rebrightening were the same between the two superoutbursts 
in EG Cnc.  
This phenomenon strengthens the interpretation 
proposed by \citet{kim16alcom}, 
that is, the rebrightening pattern is inherent to each object.  
The rebrightening type would be independent of 
the amount of the initial disk mass at the beginning of 
superoutbursts.  
This idea is consistent with the picture that the rebrightening 
type is related to the evolutionary stage of DNe, which was 
suggested by \citet{kat15wzsge}.  
\citet{kat15wzsge} also noticed that WZ Sge showed 
the same type of rebrightening in its superoutbursts 
despite a little decrease in the duration of the plateau 
stage of the main superoutburst, and other WZ Sge-type DNe 
entered superoutbursts more than twice exhibited the same 
type of rebrightening every superoutburst 
\citep{Pdot,pav12ezlyn,Pdot7}.  
This agrees with the above interpretation.  

The physical mechanism of rebrightenings observed in WZ Sge 
stars is unclear.  
As mentioned in the introduction, two possibilities were 
proposed as the trigger of rebrightening up to date: 
the enhancement of the mass transfer by irradiation 
\citep{ham00DNirradiationreview,pat02wzsge} 
and the mass reservoir outside the 3:1 resonance radius 
\citep{kat98super,hel01eruma}.  
As for the multiple rebrightenings observed in EG Cnc, 
we have detected superhumps during the rebrightening phase of 
the 2018 outburst of EG Cnc, and the $B-I$ and $J-K_{\rm S}$ 
colors were redder than those in the typical quiescence 
(see Sec.~3.3).  
This may suggest that substantial mass was left 
at the outermost region in the disk, which supplies 
some materials to the inner disk, 
as many other optical and NIR observations suggested in the past 
\citep{kat97egcnc,pat98egcnc,uem08j1021,iso15ezlyn,nak14j0754j2304,tam20j2104}.  
Also, the redder color of $B-I$ at the beginning of 
each rebrightening may indicate the disk expansion 
(see also Figure \ref{ub-bi}).  
This contradicts the enhanced mass transfer hypothesis 
since in that scenario the influx of low specific angular 
momentum material from the secondary makes the disk 
shrink at the onset of outbursts, which is reproduced by 
numerical simulations \citep{ich92diskradius}.  
The bluer $i-z$ and $J-K_{\rm S}$ colors during the rebrightening 
phase may indicate the depletion of the mass reservoir by accretion.  
We thus have found positive evidence for the mass reservoir model, 
while we have not for the enhanced mass-transfer model.  

If the disk possesses a lot of cool material even after 
the main superoutburst, the small increase of 
the viscosity could reproduce sustained rebrightenings 
as observed in EG Cnc, which was proposed by \citet{osa01egcnc}.  
We have confirmed that the optical flux in the quiescence between 
rebrightenings in the 2018 superoutburst of EG Cnc was higher 
than that in the quiescence before the main outburst and after 
the end of the rebrightening phase (see Figure \ref{overall}).  
Since more energy should be dissipated due to more efficient 
angular-momentum transfer, if the viscosity increases 
in the disk, the increase of the optical flux may represent 
the temporary enhancement of the viscosity of the disk.  
\citet{osa01egcnc} considered that the temporary enhancement of 
the viscosity is caused by MHD turbulence.  We also consider 
that the strong tidal dissipation lasting even after 
the main superoutburst could contribute to the enhancement 
as mentioned in Sec.~4.1.  

Recent theoretical work by \citet{mey15suumareb} suggested 
that the enhancement of the viscosity halts the cooling-wave 
propagation over the whole disk at the decline from 
the plateau stage of the main superoutburst, and 
repetitive reflections of the heating and cooling waves 
in the disk could be the source of the multiple 
rebrightenings.  
The heating and cooling waves are the transition waves 
propagating over the disk.  The former occurs when 
a region goes up to the hot state and the latter occurs 
when a region goes down to the cool state.  
They considered that the inner disk is permanently hot 
and that the entire disk easily jumps between the hot state and 
the cool state by the repetitive reflections.  

However, we have found that the $U-B$ color became redder 
during the rebrightening phase in the case of 2018 superoutburst 
in EG Cnc (see the top panel of Figure \ref{multi}).  This would 
mean that the inner disk became cooler during the rebrightening 
stage (see Sec.~3.3).  
Also, the $U$-band light curve is similar to the optical 
light curve.  
These imply that the heating and cooling waves 
would go back and forth not only at the outermost disk 
but also at the inner disk at some point in  
the rebrightening phase.  
On the other hand, the $B-I$ color keeps a similar level 
during quiescence between multiple rebrightenings 
(see also Figure \ref{overall-color}).  
Also, the rapid rise of the rebrightenings 
except for the last one suggests that most of the rebrightenings 
were triggered at the outer disk, i.e., outside-in outbursts.  
If the heating wave propagates outwards from the inner disk, 
we should always observe a slow rise at the beginning of 
each rebrightening, i.e., inside-out outbursts 
(see e.g., \cite{min85DNDI} for the difference between outside-in 
and inside-out outbursts).  
Taking into account our results, the inner portion of the disk 
is unlikely permanently hot in EG Cnc 
and the heating wave would be newly triggered at the beginning of 
each rebrightening at the outer disk rather than repetitive 
reflections.

\subsection{Color variations of superhumps}

We have investigated the time evolution of the color variation of 
superhumps in Sec.~3.3.  
The redder color at the superhump maximum during stage B was 
also confirmed in the other two WZ Sge stars 
\citep{iso15ezlyn,neu17j1222}, although the different types of 
color variations were found in V455 And \citep{mat09v455and}.  
\citet{ima18hvvir} showed the color variations of stage B 
superhumps change with time in the two WZ Sge stars 
HV Vir and OT J012059.6$+$325545, and interpreted that 
the time-varying pressure effect working on the precessing disk 
\citep{lub92SH,hir93SHperiod} makes the variety.  
The difference in the color variation of superhumps 
among WZ Sge-type stars maybe because of the time variation of 
the pressure effect and/or the difference of the color bands 
used for the analyses.  
We have explored the $V-R$ color, while \citet{mat09v455and} 
investigated the $V-J$ color, and the data did not cover 
the entire period of the stage B superhumps.  
On the other hand, we have confirmed the bluer color around 
the superhump maximum in the post-superoutburst stage.  
This trend was confirmed also in SSS J122221.7$-$311525 
\citep{neu17j1222}.  

The color variation of superhumps seems to be weaker 
after the main superoutburst (see also Figure \ref{SHcolor}).  
Superhumps are regarded to be produced by 
periodic tidal stressing working on the eccentric disk 
\citep{whi88tidal,hir90SHexcess,lub91SHa}.  
The difference in the amplitude of the color variation is 
likely attributed to the difference in the size of the region 
in which the eccentricity works.  
The wider region the eccentricity propagates over, 
the larger the change in the temperature, geometry, and/or 
density of the disk during each cycle of superhumps would be.  
The superhump period is the synodic period between 
the period of the prograde precession of the disk and 
the orbital period \citep{whi88tidal,hir90SHexcess,lub91SHa}, 
and the period change during stage B is believed to 
represent that the eccentricity propagates over 
the entire disk \citep{kat13qfromstageA}.  
On the other hand, the period variation during the rebrightening 
phase would be moderate in comparison with that during stage B, 
since the late-stage superhumps included two constant-period stages.

\section{Summary}

We have analyzed our multi-wavelength photometric data 
during the 2018 superoutburst of the WZ Sge-type DN EG Cnc 
to investigate whether EG Cnc is a possible candidate 
for the period bouncer and the origin of its multiple 
rebrightenings.  
The main results are as follows.  
\begin{itemize}
\item 
The rebrightenings in the 2018 superoutburst of EG Cnc 
almost completely reproduced those in its previous superoutburst 
in 1996--1997.  EG Cnc is the second object following AL Com, 
in which this kind of phenomenon was confirmed.  This phenomenon 
implies that some parameters inherent to each object are 
related to the rebrightening type in WZ Sge-type stars.  
\item The binary mass ratio of EG Cnc is estimated to be 0.045(5) 
and the possible range is estimated to be 0.019--0.057  
by using the period of the stage A superhumps and the late-stage 
superhumps that we have detected.  
Also, the amplitude of superhumps in the main superoutburst and 
the fading rate at the plateau stage where stage B superhumps 
were observed was small, which is common in period-bouncer 
candidates among WZ Sge stars.  
Although the interval between superoutbursts and the duration of 
stage A superhumps are short in this object in comparison with 
other candidates, the increase of the viscosity in the disk, 
which might be caused by the long-lasting elliptical disk 
suggested by the sustained late-stage superhumps, could affect 
at least the short recurrence time of superoutbursts.  
Taking into account the estimated mass ratio and some observational 
features common in period-bouncer candidates, we consider that 
EG Cnc is a possible candidate for the period bouncer.  
\item The $B-I$ and $J-K_{\rm S}$ colors were unusually red 
at the beginning of the rebrightening phase, and gradually became 
bluer in quiescence between rebrightenings with time.  
This would mean that cool material existed at the outermost disk 
before multiple rebrightenings and was depleted during 
the rebrightening phase.  
The slightly redder $B-I$ color at the rising part of each rebrightening 
suggests the disk expansion.  
These results support the idea of the mass reservoir as the mechanism 
of rebrightenings rather than the enhancement of the mass-transfer rate.  
\item
The $UV$ light curve well traced the optical light variations 
of multiple rebrightenings and the $U-B$ color became redder during 
the rebrightening phase, which means that the inner disk became cooler.  
The thermal-viscous instability may have newly occurred at each rebrightening.  
\end{itemize}

\section*{Acknowledgements}

This work was financially supported by Grants-in-Aid for JSPS Fellows
for young researchers (M.~Kimura).  
M.~Kimura acknowledges support by the Special Postdoctoral Researchers
Program at RIKEN.  
We are thankful to many amateur observers for providing a lot of 
the data used in this research.  
This work is also supported by the Optical and Near-infrared Astronomy 
Inter-University Cooperation Program.  
M.~Kimura is grateful to K.~Takagi, I.~Otsubo, Y.~Yamazaki, and H.~Kimura 
at Higashi-Hiroshima Observatory and Y.~Uzawa at Saitama University 
for engaging in the observation of this star.  
The work of N.~Katysheva and S.~Shugarov was supported by the Program 
of Development of Lomonosov Moscow State University `Leading 
Scientific Schools', project `Physics of Stars, Relativistic 
Objects and Galaxies'. 
The work by S.~Shugarov was supported by the Slovak Research and 
Development Agency under the contract No. APVV-15-0458 and by 
the Slovak Academy of Sciences grant VEGA No.~2/0008/17.
The work by P.~A. Dubovsky, I.~Kudzej, and S.~Shugarov was supported 
by the Slovak Research and Development Agency under the contract 
No.~APVV-15-0458.  
We thank the anonymous referee for helpful comments.

\section*{Supporting information}

Additional supporting information can be found in the online version 
of this article:
Supplementary tables 1 and 2.
Supplementary data is available at PASJ Journal online.   


\newcommand{\noop}[1]{}

\end{document}